\documentclass[]{pasj00}
\SetRunningHead{H. Uchiyama et al.}
{Global Distribution of Fe K$\alpha$ Lines in the Galactic Center Region}

\begin{document}
\title{Global Distribution of Fe K$\alpha$ Lines in the Galactic Center Region Observed with the Suzaku Satellite}
\author{Hideki \textsc{Uchiyama}}
\affil{Department of Physics, Graduate School of Science, The University of Tokyo, 7-3-1 Hongo, Bunkyo-ku, Tokyo 113-0033}
\email{uchiyama@juno.phys.s.u-tokyo.ac.jp}
\author{Masayoshi \textsc{Nobukawa}, Takeshi Go \textsc{Tsuru}, Katsuji \textsc{Koyama}} 
\affil{Department of Physics, Graduate school of Science, Kyoto 
University, Oiwake-cho, Kitashirakawa, Kyoto 606-8502}
\and
\author{Hironori \textsc{Matsumoto}}
\affil{Kobayashi-Maskawa Institute for the Origin of Particles and the Universe, Nagoya University, Furo-cho, Chikusa-ku, Nagoya, 464-8602}

\KeyWords{Galaxy: Center---Inter stellar medium--- X-ray spectra}
\maketitle

\begin{abstract}

We have surveyed spatial profiles of the Fe K$\alpha$ lines in the Galactic center 
diffuse X-rays (GCDX), including the transient region from the GCDX to the Galactic ridge 
X-ray emission (GRXE), with the Suzaku satellite. 
We resolved Fe K$\alpha$ line complex into three lines of Fe \emissiontype{I}, 
Fe \emissiontype{XXV} and Fe \emissiontype{XXVI} K$\alpha$, 
and obtained their spatial intensity profiles 
with the resolution of $\sim \timeform{0D.1}$.
We compared the Fe \emissiontype{XXV} K$\alpha$ profile
with a stellar mass distribution (SMD) model made from near infrared observations.
The intensity profile of Fe \emissiontype{XXV} K$\alpha$ is nicely fitted with the SMD 
model in the GRXE region, while that in the GCDX region shows 
$3.8\pm0.3$ $(\timeform{0D.2}<|l|<\timeform{1D.5})$ or $19\pm6$ $(|l|<\timeform{0D.2})$ times excess 
over the best-fit SMD model in the GRXE region. 
Thus Fe \emissiontype{XXV} K$\alpha$ in the GCDX is
hardly explained by the same origin of the GRXE. 
In the case of point source origin, a new population 
with the extremely strong Fe \emissiontype{XXV} K$\alpha$ line is required.  
An alternative possibility is that the majority of the GCDX is truly diffuse optically thin
thermal plasma.

\end{abstract}

\section{Introduction} 
Diffuse X-rays were discovered from the inner disk of the Milky Way Galaxy 
with the HEAO-1 satellite \citep{Wo82}, and called as the Galactic ridge 
X-ray emission (GRXE).  EXOSAT observed more details and found that the 
GRXE were extended by about $\pm \timeform{60D}$ in longitude and $\pm (\timeform{1D}$--$\timeform{2D}$)
in latitude \citep{Wa85}. The Tenma satellite discovered strong emission 
lines at the energy of about 6.7 keV (here the 6.7-keV line) in the GRXE \citep{Ko86}. 
The Ginga satellite made survey observations along the Galactic plane with the 6.7-keV line, 
and found a flux peak at the Galactic center with the width of about $\timeform{1D}$. 
We call this peak as the Galactic center diffuse X-rays (GCDX) \citep{Ko89, Ya90, Ya93}.

The ASCA satellite separated the 6.7-keV line in the GCDX into three emission lines at 6.4, 6.7 and 7.0 keV, 
which were respectively identified as K$\alpha$ lines from neutral (Fe \emissiontype{I}), 
He-like (Fe \emissiontype{XXV}) and H-like (Fe \emissiontype{XXVI}) iron 
\citep{Ko96}. For the GRXE, the Suzaku satellite resolved the Fe K$\alpha$ lines \citep{Eb08}.  
The strongest emission line in the GCDX and GRXE is 
Fe \emissiontype{XXV} K$\alpha$ at 6.7 keV, and hence would be due to an optically thin thermal plasma. 
From the flux ratio of Fe \emissiontype{XXVI} K$\alpha$/Fe \emissiontype{XXV} K$\alpha$, 
the plasma temperature of the GCDX was estimated to be around $kT=$5--10 keV (e.g. \cite{Ko07b, Ya09}).

Since spectra of the GCDX and GRXE have spatial temperature
variations and/or may be contaminated by possible non-thermal
emissions, we concentrate on the flux of Fe \emissiontype{XXV} K$\alpha$
line as a pure component of the thin thermal plasma. Previous
studies of the Fe K$\alpha$ lines with the Ginga and RXTE
satellites \citep{Ya93,Re06}  had  limited spatial resolutions
(FWHM $\sim \timeform{1D}$--$\timeform{2D}$)  compared to the
size of the GCDX, and the Fe \emissiontype{XXV} K$\alpha$ line flux was contaminated
by the adjacent Fe \emissiontype{I} and \emissiontype{XXVI} K$\alpha$ lines due to the
limited energy resolutions (FWHM $\sim$ 1.1 keV at 6 keV).
\citet{Ya09} reported that the intensity ratio of  the three
Fe K$\alpha$ lines has spatial dependence and hence to resolve the
three lines is essential for the study the nature of Fe \emissiontype{XXV} K$\alpha$ line.
This paper reports, for the first time, the Fe \emissiontype{XXV} K$\alpha$ line
distributions in the GCDX and the transient region from the GCDX to
the GRXE ($\timeform{-3D}< l < \timeform{2D}$, $\timeform{-2D} < b
<\timeform{1D}$) with the CCD cameras onboard the Suzaku satellite.

The observations and data reduction are described in section 2. The analysis and 
the intensities of Fe \emissiontype{I} K$\alpha$, Fe \emissiontype{XXV} K$\alpha$ 
and Fe \emissiontype{XXVI} K$\alpha$ are given in section 3. 
In section 4, we compare the Fe \emissiontype{XXV} K$\alpha$ profiles of the GCDX and GRXE 
with that of a stellar mass model made from near infrared (NIR) observations.  
Based on the profile comparison between NIR stars 
and Fe \emissiontype{XXV} K$\alpha$, we discuss origin of the GCDX.  
In this paper, uncertainties are quoted at the 90\% confidence range unless otherwise stated.

\section{Observations and Data Reduction}
The data in this paper were obtained from September 2005 to March 2009 with the 
X-ray Imaging Spectrometer (XIS) onboard the Suzaku satellite.  
Most of the data were observed as the Suzaku Key project observation.
The total effective exposure is about 3.2~Ms. 
The details of these observations are summarized in table \ref{tab:obslog_GC}. 

The XIS contains four sets of X-ray CCD camera systems (XIS~0, 1, 2, and 3) placed on 
the focal planes of four X-Ray Telescopes (XRTs) onboard the  Suzaku satellite. XIS~0, 2, and 3 have 
front-illuminated (FI) CCDs, while XIS~1 has a back-illuminated (BI) CCD. 
Detailed descriptions of the Suzaku satellite, the XRT, and the XIS can 
be found in \citet{Mi07}, \citet{Se07}, and \citet{Ko07a}, respectively.

In these observations, the XIS was operated in the normal clocking full window mode,
with the time resolution of 8.0~s.
One of the FI CCD cameras (XIS~2) has been out of function since November 2006, 
and hence 71\% of the total exposure was observed without XIS~2.   
Although the XIS CCDs were significantly degraded by on-orbit particle radiation,
the CCD performance had been calibrated well with the checker-flag charge-injection (CI) 
method \citep{Na08,Oz09} until October 2006 , and has been restored  
with the spaced-row CI (SCI) method \citep{Pr08,Uc09}. 
The SCI  method was used in 75\% of the total exposure.
The XIS~2 and SCI status in the observations is shown in table \ref{tab:obslog_GC}. 

We reprocessed the data using the calibration data base (CALDB) 
released at 2008-10-20 and {\tt xispi} in the HEADAS software package version 6.6.1. 
We confirmed that the center energy of 
Mn \emissiontype{I} K$\alpha$ line of  the onboard calibration sources ($^{55}$Fe) was 
consistent with 5.895~eV 
within the calibration uncertainty of 20 eV, and
the energy resolutions were 130--180 eV (FWHM) at 5.9 keV during the observations.

\renewcommand{\arraystretch}{0.8} 
\begin{table*}
 \caption{Observation data list. } \label{tab:obslog_GC}
\scalebox{0.7}{\begin{minipage}{\textheight}
 \begin{tabular}{ccccccccccc}
 \hline
 \hline
OBSID& \multicolumn{4}{c}{Pointing direction} & \multicolumn{2}{c}{Observation} &  Exp. & XIS\,2$^*$ & SCI$^{\dagger}$ & Object name\\
& $\alpha_{\rm 2000.0}(\timeform{D})$ & $\delta_{\rm 2000.0}(\timeform{D})$ & $l(\timeform{D})$ & $b(\timeform{D})$ & Start (UT)  & End (UT) &  (ks) & &  & \\
\hline
100027010 & 266.51& -28.92 &0.05& -0.07&  2005-09-23T07:18:25 & 2005-09-24T11:05:19 &44.7 & ok & off& Sgr A East\\
100027020 & 266.30 & -29.17 & -0.25 & -0.05 & 2005-09-24T14:17:17 & 2005-09-25T17:27:19 & 37.4 & ok & off & Sgr A west \\
100037010 & 266.30 & -29.17 & -0.25 & -0.05 & 2005-09-29T04:35:41 & 2005-09-30T04:29:19 & 43.5 & ok & off & Sgr A west \\
100037030 & 266.20 & -29.35 & -0.45 & -0.07 & 2005-09-30T06:06:32 & 2005-09-30T07:41:20 & 3.0 & ok & off & KS1741-293(GCBGD2) \\
100037040 & 266.51 & -28.93 & 0.06 & -0.07 & 2005-09-30T07:43:01 & 2005-10-01T06:21:24 & 42.9 & ok & off & Sgr A East \\
100037060 & 266.88 & -28.44 & 0.64 & -0.10 & 2005-10-10T12:28:01 & 2005-10-12T07:05:23 & 76.4 & ok & off & Sgr B2 \\
500018010 & 266.16 & -29.47 & -0.57 & -0.09 & 2006-02-20T12:45:25 & 2006-02-23T10:50:14 & 106.9 & ok & off & SGR C \\
500005010 & 266.77 & -28.63 & 0.43 & -0.12 & 2006-03-27T23:00:22 & 2006-03-29T18:12:15 & 88.4 & ok & off & GC CENTER \#2 \\
100048010 & 266.51 & -28.93 & 0.06 & -0.07 & 2006-09-08T02:23:24 & 2006-09-09T09:06:15 & 63.0 & ok & off & GALACTIC CENTER \\
501040010 & 266.69 & -28.38 & 0.61 & 0.07 & 2006-09-21T17:29:01 & 2006-09-23T06:52:20 & 61.4 & ok & on & GC SGR B NORTH \\
501040020 & 266.69 & -28.38 & 0.61 & 0.07 & 2006-09-24T05:03:12 & 2006-09-25T07:00:24 & 44.8 & ok & on & GC SGR B NORTH \\
501008010 & 266.50 & -29.17 & -0.15 & -0.19 & 2006-09-26T14:18:16 & 2006-09-29T21:25:14 & 129.6 & ok & on & GC SOUTH \\
501009010 & 266.19 & -28.91 & -0.07 & 0.18 & 2006-09-29T21:26:07 & 2006-10-01T06:55:19 & 51.2 & ok & on & GC SOUTH BGD \\
501010010 & 266.26 & -30.37 & -1.29 & -0.64 & 2006-10-07T02:16:52 & 2006-10-08T10:19:19 & 50.7 & ok & off & HESS J1745-303 \\
501049010 & 265.38 & -29.75 & -1.17 & 0.33 & 2006-10-08T10:22:40 & 2006-10-09T02:19:24 & 19.6 & ok & on & GALACTIC CENTER \\
501050010 & 265.91 & -29.65 & -0.83 & 0.00 & 2006-10-09T02:20:25 & 2006-10-09T13:39:24 & 22.0 & ok & on & GALACTIC CENTER \\
501051010 & 265.70 & -29.93 & -1.17 & -0.00 & 2006-10-09T13:40:09 & 2006-10-10T06:44:24 & 21.9 & ok & on & GALACTIC CENTER \\
501052010 & 265.50 & -30.21 & -1.50 & 0.00 & 2006-10-10T06:45:09 & 2006-10-10T21:18:14 & 19.3 & ok & on & GALACTIC CENTER \\
501053010 & 265.30 & -30.50 & -1.83 & -0.00 & 2006-10-10T21:18:59 & 2006-10-11T10:06:14 & 21.9 & ok & on & GALACTIC CENTER \\
501057010 & 266.03 & -30.11 & -1.17 & -0.33 & 2006-10-11T10:07:27 & 2006-10-12T03:28:14 & 20.5 & ok & on & GALACTIC CENTER \\
501039010 & 267.02 & -28.35 & 0.78 & -0.16 & 2007-03-03T12:20:20 & 2007-03-05T12:39:25 & 96.4 & & on & GC SGR B EAST \\
501046010 & 265.98 & -28.90 & -0.17 & 0.33 & 2007-03-10T15:03:10 & 2007-03-11T03:55:14 & 25.2 & & on & GALACTIC CENTER \\
501047010 & 265.78 & -29.19 & -0.50 & 0.33 & 2007-03-11T03:55:59 & 2007-03-11T19:04:14 & 25.6 & & on & GALACTIC CENTER GC2 \\
501048010 & 265.58 & -29.47 & -0.83 & 0.33 & 2007-03-11T19:04:59 & 2007-03-12T08:09:14 & 27.5 & & on & GALACTIC CENTER GC3 \\
501054010 & 266.63 & -29.25 & -0.17 & -0.33 & 2007-03-12T08:11:07 & 2007-03-12T23:58:24 & 26.1 & & on & GALACTIC CENTER GC9 \\
501055010 & 266.43 & -29.54 & -0.50 & -0.33 & 2007-03-12T23:59:09 & 2007-03-13T15:40:19 & 27.2 & & on & GALACTIC CENTER GC10 \\
501056010 & 266.23 & -29.82 & -0.83 & -0.33 & 2007-03-13T15:41:12 & 2007-03-14T05:00:24 & 26.5 & & on & GALACTIC CENTER GC11 \\
501058010 & 266.98 & -27.72 & 1.30 & 0.20 & 2007-03-14T05:02:29 & 2007-03-15T18:55:14 & 63.3 & & on & GC SGR D NORTH \\
501059010 & 267.09 & -27.94 & 1.17 & 0.00 & 2007-03-15T18:55:51 & 2007-03-17T05:06:19 & 62.2 & & on & GC SGR D \\
501060010 & 267.29 & -27.65 & 1.50 & 0.00 & 2007-03-17T05:07:04 & 2007-03-18T20:58:14 & 64.8 & & on & GC SGR D EAST \\
502022010 & 266.81 & -28.88 & 0.23 & -0.27 & 2007-08-31T12:33:33 & 2007-09-03T19:00:25 & 134.8 & & on &  (L,B)=(0.25,-0.27) \\ 
102013010 & 266.51 & -28.93 & 0.06 & -0.07 & 2007-09-03T19:01:10 & 2007-09-05T05:20:20 & 51.4 & & on & GALACTIC CENTER \\
502020010 & 267.19 & -28.13 & 1.05 & -0.17 & 2007-09-06T00:26:47 & 2007-09-09T16:10:19 & 139.1 & & on & SGR D SNR \\
502002010 & 267.16 & -29.14 & 0.17 & -0.67 & 2007-10-09T16:40:54 & 2007-10-10T03:40:24 & 23.2 & & on & GC14 \\
502003010 & 266.96 & -29.42 & -0.17 & -0.67 & 2007-10-10T03:41:13 & 2007-10-10T15:20:24 & 21.5 & & on & GC15 \\
502004010 & 267.48 & -29.31 & 0.17 & -1.00 & 2007-10-10T15:21:17 & 2007-10-11T01:00:24 & 19.9 & & on & GC16 \\
502005010 & 267.29 & -29.60 & -0.17 & -1.00 & 2007-10-11T01:01:17 & 2007-10-11T11:32:20 & 20.6 & & on & GC17 \\
502006010 & 266.18 & -28.62 & 0.17 & 0.33 & 2007-10-11T11:34:01 & 2007-10-11T23:07:14 & 22.6 & & on & GC18 \\
502007010 & 265.86 & -28.45 & 0.17 & 0.67 & 2007-10-11T23:09:15 & 2007-10-12T09:52:14 & 22.0 & & on & GC19 \\
502008010 & 265.66 & -28.73 & -0.17 & 0.67 & 2007-10-12T09:52:59 & 2007-10-12T21:50:19 & 23.8 & & on & GC20 \\
502009010 & 267.48 & -27.37 & 1.83 & -0.00 & 2007-10-12T21:52:24 & 2007-10-13T07:30:19 & 20.9 & & on & GC21 \\
502010010 & 266.38 & -28.34 & 0.50 & 0.33 & 2007-10-13T07:32:00 & 2007-10-13T18:50:24 & 21.6 & & on & GC22 \\
502011010 & 266.57 & -28.05 & 0.83 & 0.33 & 2007-10-13T18:51:09 & 2007-10-14T05:30:24 & 23.0 & & on & GC23 \\
502016010 & 266.23 & -30.11 & -1.08 & -0.48 & 2008-03-02T18:08:00 & 2008-03-04T17:40:19 & 70.5 & & on & HESS J1745-303 1 \\
502017010 & 266.47 & -30.09 & -0.95 & -0.65 & 2008-03-06T13:26:36 & 2008-03-08T16:00:24 & 72.6 & & on & HESS J1745-303 2 \\
502018010 & 266.06 & -30.24 & -1.27 & -0.43 & 2008-03-08T16:02:17 & 2008-03-10T21:00:19 & 79.0 & & on & HESS J1745-303 3 \\
502051010 & 266.94 & -28.15 & 0.91 & 0.01 & 2008-03-11T06:19:45 & 2008-03-15T05:30:18 & 138.8 & & on & GC G0.9+0.1 \\
502059010 & 268.38 & -29.96 & -0.00 & -2.00 & 2007-09-29T01:40:51 & 2007-10-02T14:10:16 & 136.8 & & on & GALACTIC BULGE \\
503007010 & 266.44 & -28.57 & 0.33 & 0.17 & 2008-09-02T10:15:27 & 2008-09-03T22:52:24 & 52.2 & & on & GC LARGEPROJECT1 \\
503008010 & 266.78 & -29.13 & 0.00 & -0.38 & 2008-09-03T22:53:29 & 2008-09-05T06:56:19 & 53.7 & & on & GC LARGEPROJECT2 \\
503009010 & 266.45 & -29.34 & -0.33 & -0.24 & 2008-09-05T06:57:08 & 2008-09-06T15:55:24 & 52.4 & & on & GC LARGEPROJECT3 \\
503010010 & 266.04 & -29.55 & -0.70 & -0.05 & 2008-09-06T15:56:13 & 2008-09-08T01:39:24 & 53.1 & & on & GC LARGEPROJECT4 \\
503011010 & 265.95 & -29.83 & -0.97 & -0.13 & 2008-09-08T09:08:09 & 2008-09-09T21:33:19 & 57.6 & & on & GC LARGEPROJECT5 \\
503012010 & 266.30 & -29.94 & -0.91 & -0.44 & 2008-09-14T19:35:07 & 2008-09-16T00:50:14 & 57.7 & & on & GC LARGEPROJECT6 \\
503013010 & 265.67 & -30.07 & -1.30 & -0.05 & 2008-09-16T00:51:19 & 2008-09-18T04:44:24 & 104.8 & & on & GC LARGEPROJECT7 \\
503014010 & 265.18 & -30.75 & -2.10 & -0.05 & 2008-09-18T04:46:49 & 2008-09-19T07:32:20 & 55.4 & & on & GC LARGEPROJECT8 \\
503015010 & 265.03 & -30.96 & -2.35 & -0.05 & 2008-09-19T07:33:05 & 2008-09-20T09:56:13 & 56.8 & & on & GC LARGEPROJECT9 \\
503016010 & 264.87 & -31.17 & -2.60 & -0.05 & 2008-09-22T06:47:49 & 2008-09-23T08:07:17 & 52.2 & & on & GC LARGEPROJECT10 \\
503017010 & 264.72 & -31.38 & -2.85 & -0.05 & 2008-09-23T08:08:10 & 2008-09-24T09:21:13 & 51.3 & & on & GC LARGEPROJECT11 \\
503018010 & 264.56 & -31.60 & -3.10 & -0.05 & 2008-09-24T09:27:54 & 2008-09-24T22:30:24 & 29.4 & & on & GC LARGEPROJECT12 \\
503072010 & 265.99 & -29.21 & -0.42 & 0.17 & 2009-03-06T02:39:11 & 2009-03-09T02:55:24 & 140.6 & & on & EXTENDED CHIMNEY \\
403009010 & 266.48 & -28.78 & 0.17 & 0.02 & 2009-03-21T02:03:27 & 2009-03-23T18:26:09 & 110.8 & & on & ARCHES CLUSTER \\
\hline
\end{tabular}
$^*$The XIS2 column shows whether the XIS~2 worked (``ok'') or not in the observation. \\
$^{\dagger}$The SCI column shows whether the SCI mode was on or off in the observation.
\end{minipage}}
\end{table*}
 \renewcommand{\arraystretch}{1.0}

\section{Analysis and Results}
\begin{figure}[hbtp]
		\begin{center}
		(a) \FigureFile(80mm,60mm){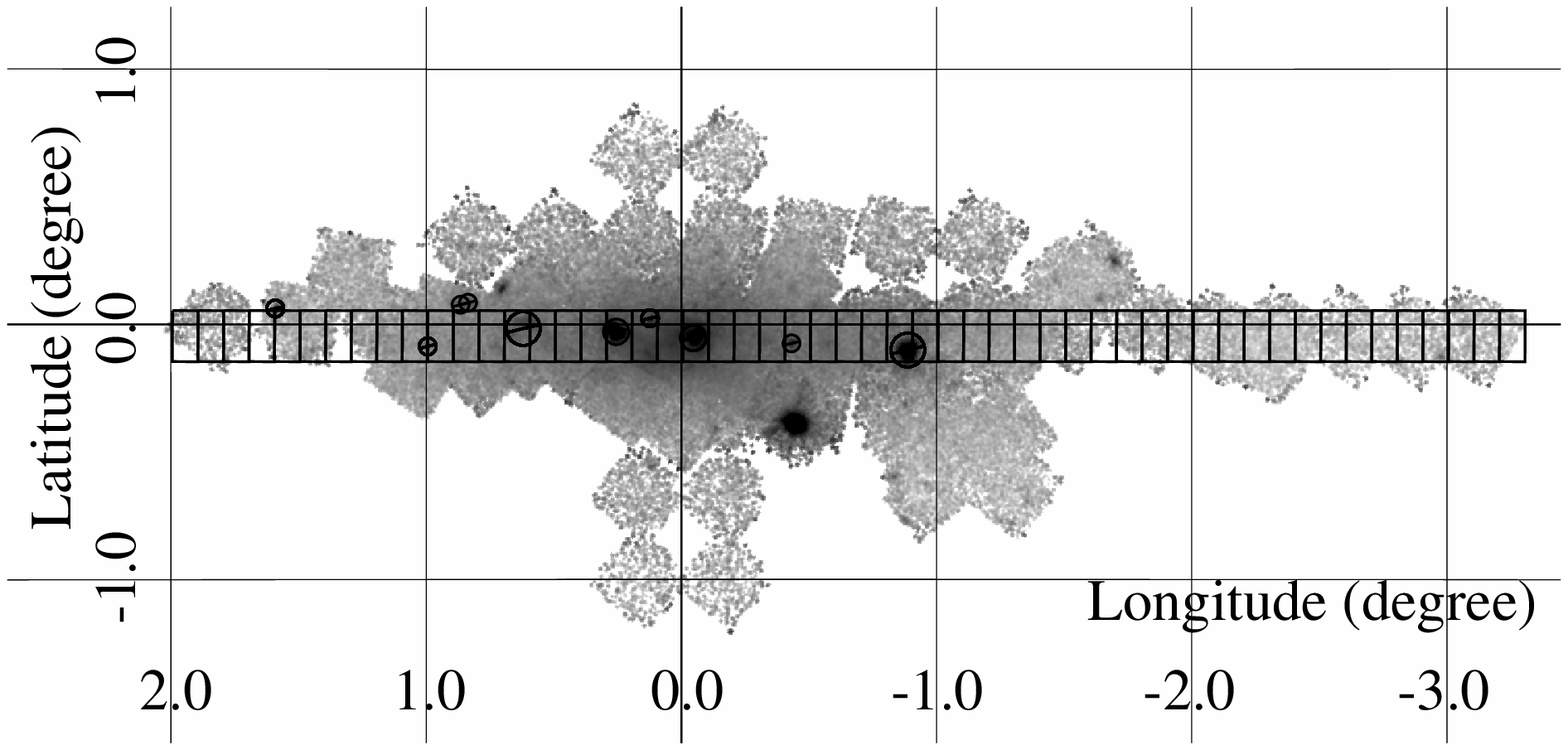}\\
		(b) \FigureFile(80mm,60mm){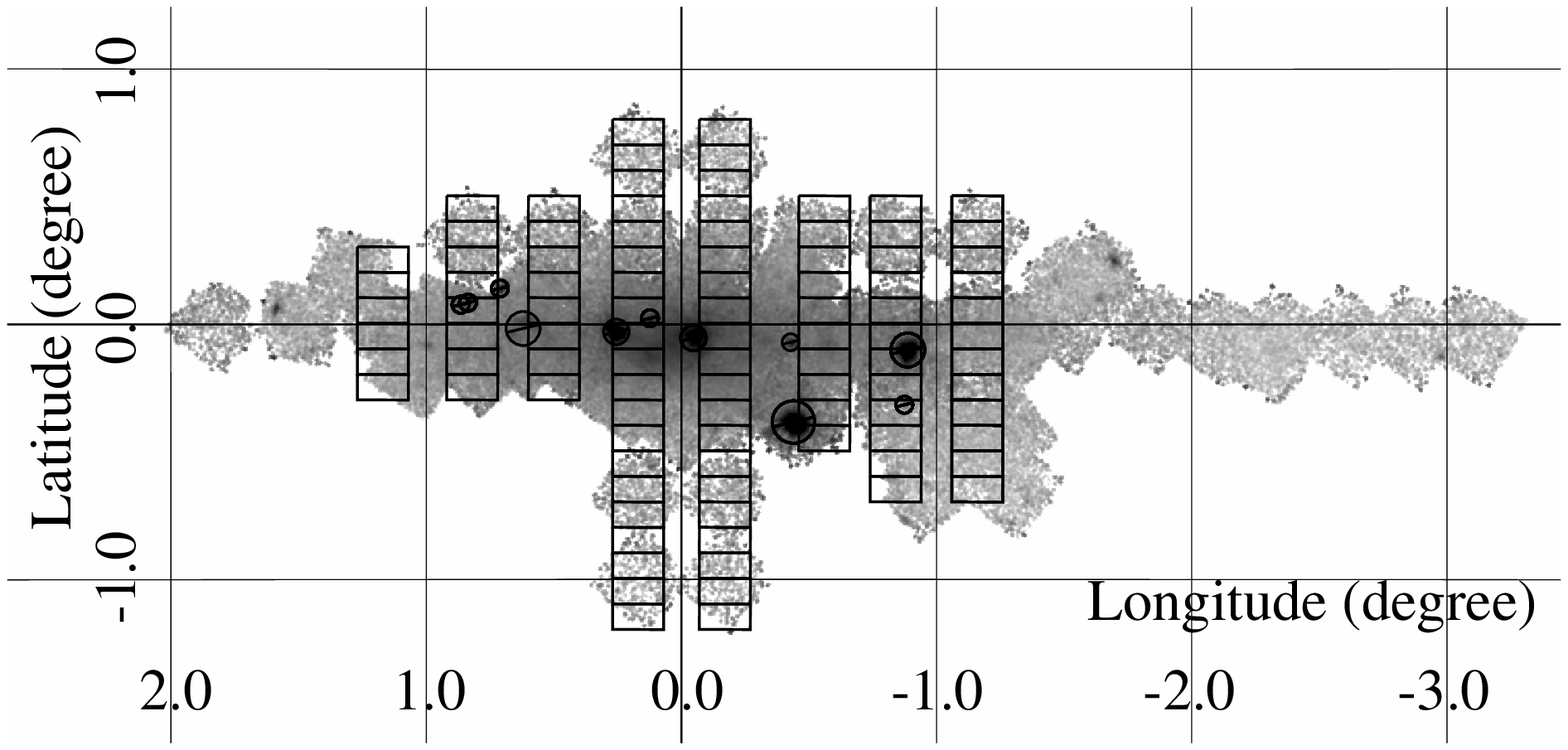}
		\end{center}
		\caption{(a) The rectangles to extract X-ray spectra for the longitudinal distribution overlaid 
		on the Suzaku XIS image in the 6.55--6.8 keV band.  Bright point source regions (circles, see table
		 \ref{tab:brsrc}) are excluded before making the spectra. (b) Same as (a) but for the latitudinal distribution.}
		\label{fig:Fig1}
\end{figure}

In order to investigate  intensity profiles of Fe K$\alpha$ lines along
the Galactic longitude and latitude, X-ray spectra were made from the
rectangles shown in figures \ref{fig:Fig1}a and b, respectively. 
The sizes of the rectangles along the longitude (figure \ref{fig:Fig1}a) are $\timeform{6'} \times \timeform{12'}$ 
($\timeform{0D.1}$ in longitude and $\timeform{0D.2}$  in latitude), 
while those along the latitude (figure \ref{fig:Fig1}b) are $\timeform{12'} \times 
\timeform{6'}$ ($\timeform{0D.2}$ in longitude 
and $\timeform{0D.1}$  in latitude).
Here we adopted a new coordinate as  $(l_*, b_*)=(l+\timeform{0D.056}$, $b+\timeform{0D.046}$),  
where the origin of this coordinate $(l_*, b_*)=(\timeform{0D}, \timeform{0D})$ 
is the position of Sgr A$^*$ \citep{Re04} and the $b_*=\timeform{0D}$ line corresponds the Galactic plane.
With this new coordinate, the centers of each rectangle along the longitude (figure \ref{fig:Fig1}a) are given as 
$(l_*, b_*)= (\timeform{0D.1}\times n$, $\timeform{0D})$ where $n$ is an integer number. 
The centers of each rectangle along the latitude (figure \ref{fig:Fig1}b) are $b_*=\timeform{0D.1} \times n$
and $l_*= \timeform{-1D.104}$, $\timeform{-0D.784}$, $\timeform{-0D.504}$, $\timeform{-0D.114}$, $\timeform{0D.226}$, 
$\timeform{0D.556}$, $\timeform{0D.876}$ and $\timeform{1D.226}$. 
To extend the profile to larger $b_*$ values, we used the data at $(l_*, b_*)=(\timeform{0D.056}$, $\timeform{-1D.954})$ (OBSID 502059010). 
Since the Fe K$\alpha$ flux at this position is lower than any other regions, 
we made a spectrum of  $\timeform{17'} \times \timeform{17'}$ rectangle.
To make spectra of diffuse emission, bright sources listed in table \ref{tab:brsrc}  and 
the CCD corners illuminated by the calibration sources were excluded from all the rectangles.
 
Using {\tt xissimarfgen} \citep{Is07} and {\tt xisrmfgen}, the effective area of the XRT 
and the response of the XIS  were calculated, and ancillary response files 
and redistribution matrix files were made for each observation.
Since the responses of the FIs  are almost the same, we merged the FI spectra. 

The non X-ray background (NXB) spectra sorted by the cut-off rigidity  were made from the same 
source region in the detector coordinate using {\tt xisnxbgen} \citep{Ta08}.
The spectra of NXB were subtracted from those of the GCDX.

\begin{table*}
	\caption{The positions and radii of bright X-ray sources excluded before making the diffuse X-ray spectra.}
	\label{tab:brsrc}
	\begin{center}
    	\begin{tabular}{crrc}
\hline
\hline
Name &\multicolumn{2}{c}{Position}& Radius  \\
& \multicolumn{1}{c}{$l$} &\multicolumn{1}{c}{$b$} & \\
\hline
1E 1740.7$-$2942 		&$\timeform{-0D.887}$& $\timeform{-0D.097}$& \timeform{4'}\\ 
CXOGCS J174445.5$-$295042 &$\timeform{-0D.774}$& $\timeform{-0D.062}$& \timeform{2'}\\
2E~1742.9$-$2929	&$\timeform{-0D.439}$& $\timeform{-0D.383}$& \timeform{5'}\\
KS~1741$-$293 		&$\timeform{-0D.432}$& $\timeform{-0D.074}$& \timeform{3'} \\ 
Sgr A East				&$\timeform{-0D.047}$& $\timeform{-0D.053}$& \timeform{3'}\\ 
Arches cluster 		&$\timeform{ 0D.123}$& $\timeform{0D.024}$& \timeform{2'}\\ 
1E~1743.1$-$2843 	&$\timeform{0D.256}$& $\timeform{-0D.029}$& \timeform{3'}\\ 
G0.61+0.01 			&$\timeform{0D.620}$& $\timeform{0D.016}$& \timeform{4'}\\ 
CXOGC J174645.3$-$281546 &$\timeform{0D.712}$& $ \timeform{0D.141}$& \timeform{2'}\\ 
HD 161507 			&$\timeform{0D.837}$& $\timeform{0D.085}$& \timeform{2'}\\ 
G0.9+0.1 				&$\timeform{0D.866}$& $\timeform{0D.077}$& \timeform{2'}\\ 
SAX J1748.2$-$2808  &$\timeform{0D.994}$& $\timeform{-0D.086}$& \timeform{2'}\\ 
AX J1749.1$-$2733 	&$\timeform{1D.592}$& $\timeform{ -0D.062} $&\timeform{2'}\\  
\hline
\end{tabular}
\end{center}
\end{table*}

We fitted the spectra  in the 5--10 keV band with a phenomenological model which consists of 
four Gaussians and an absorbed power-law continuum. 
The line centers of the Gaussians were fixed to 6.40 (Fe \emissiontype{I} K$\alpha$), 6.68 (Fe \emissiontype{XXV} K$\alpha$), 
6.97 (Fe \emissiontype {XXVI} K$\alpha$) and 7.06 (Fe \emissiontype{I} K$\beta$)~keV according to 
the APEC model \citep{Sm01} and \citet{Ka93}. The intrinsic widths of these lines were fixed to 0~eV. 
The intensities of the Fe \emissiontype{I} 
K$\alpha$, Fe \emissiontype{XXV} K$\alpha$, and Fe \emissiontype
{XXVI} K$\alpha$ lines in the model were free parameters. The intensity of the Fe \emissiontype{I} 
K$\beta$ line was fixed to 0.125 times that of the Fe \emissiontype{I} K$\alpha$ line
according to \citet{Ka93}.
The normalization, photon index, and interstellar absorption of the absorbed power-law model were free.
The cross section of the photoelectric absorption was obtained from \citet{Mo83}.

The FIs and BI spectra were fitted simultaneously.
We did not use the BI data above 8 keV, because the NXB fluxes become high in this energy band.
If one rectangle region overlaps more than two separate observations, we made separate spectra and were fitted 
simultaneously.

\begin{figure*}[htbp]
		\begin{center}
		\FigureFile(80mm,60mm){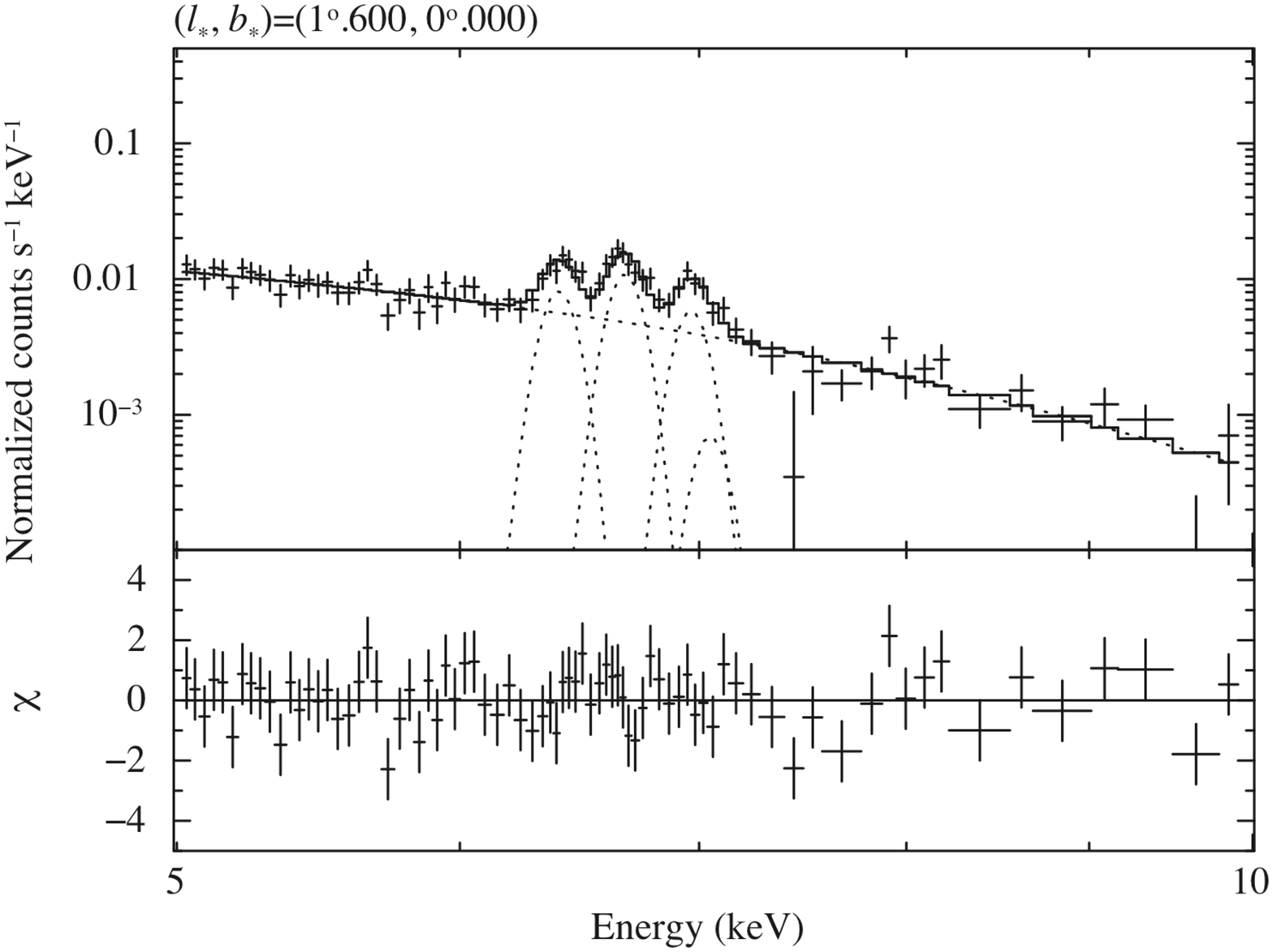}
		\FigureFile(80mm,60mm){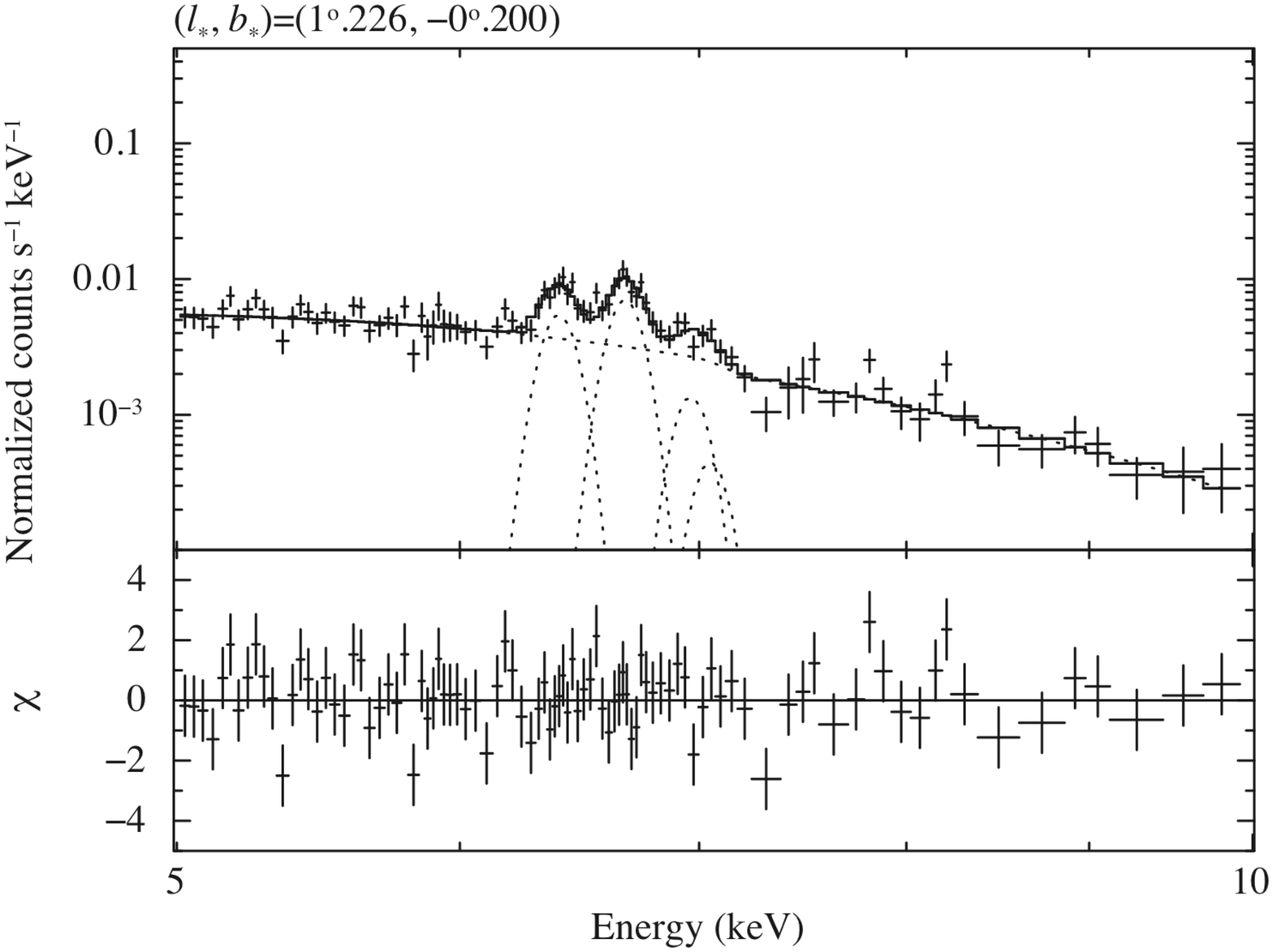}		
		\end{center}
		\caption{Left panel: an example of the spectra for the longitudinal distribution with the best-fit model
                (solid line). Dotted Gaussians are  K$\alpha$ lines of  Fe \emissiontype{I}, Fe \emissiontype{XXV} and 
		Fe \emissiontype{XXVI}.
                 The weak line in the right is  Fe \emissiontype{I} K$\beta$. For simplicity, only the FIs spectra are shown. 
		The vertical bars  show the 1$\sigma$ errors. Residuals from the best-fit model near at 8--9 keV are due to
                Ni \emissiontype{XXVII} K$\alpha$ and Fe \emissiontype{XXV} K$\beta$ and other weak lines (see text).
		 Right panel: the same as the left panel, but that for the latitudinal distribution data.
		 }
		\label{fig:SpectraExamples}
\end{figure*}

\begin{figure*}[htbp]
		\begin{center}
		(a)\FigureFile(80mm,60mm){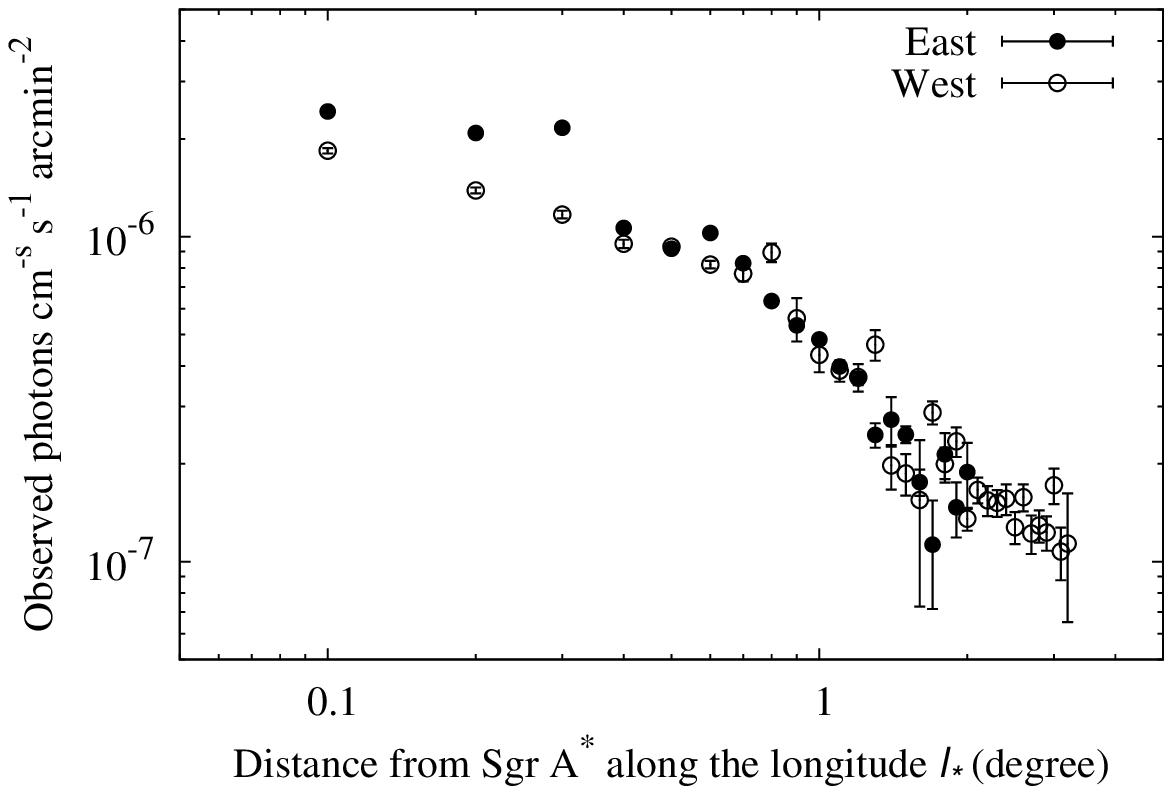}
		(b)\FigureFile(80mm,60mm){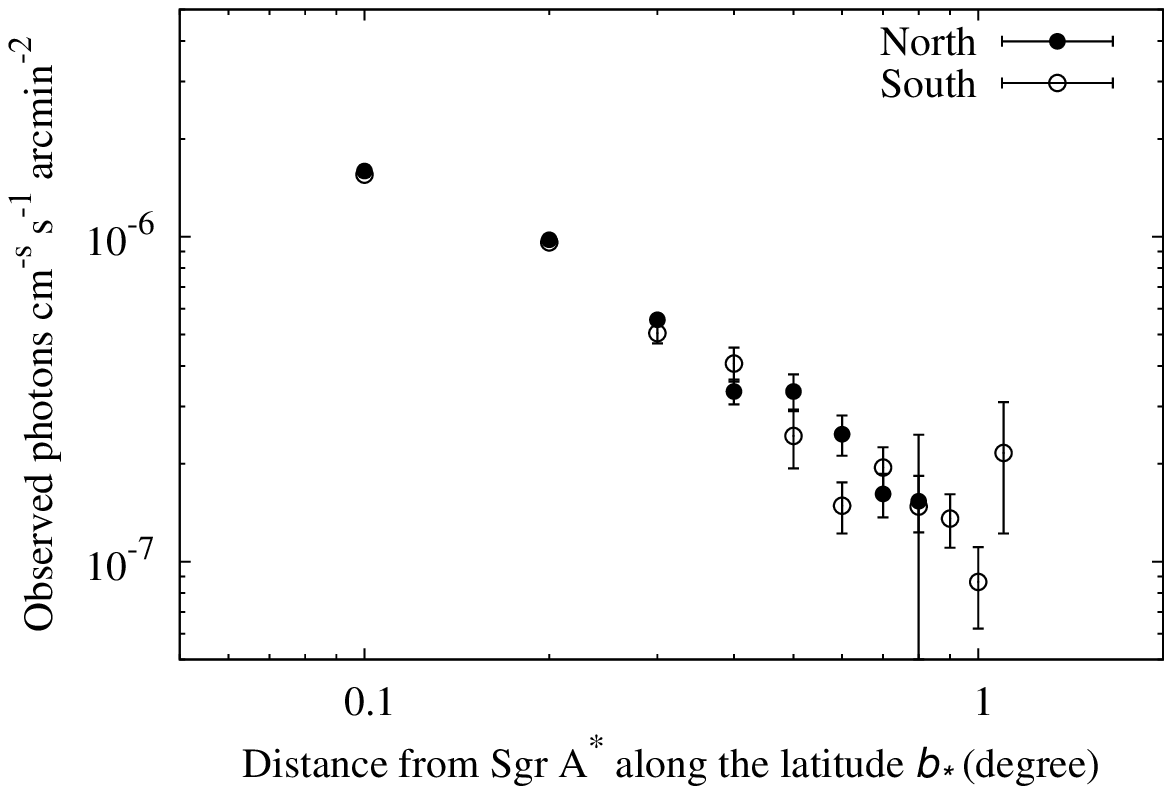} 
		\end{center}	
		\caption{(a) Observed intensity profiles of the Fe \emissiontype{XXV} K$\alpha$
		along the Galactic longitude ($b_*=\timeform{0D}$). The bars show the 1$\sigma$ errors.  
                   (b) The same as (a) but for the latitude distribution at $l_*=\timeform{-0D.114}$. }
\label{fig:FeHeKaProfile}
\end{figure*}

Some examples of the spectra and the best-fit models are shown in figure \ref{fig:SpectraExamples}. 
In these figures, we find some residuals at 8--9 keV band, which are identified as Fe \emissiontype{XXV} K$\beta$, 
Ni \emissiontype{XXVII} K$\alpha$ and other weak lines \citep{Ko07b}.  Inclusion of these lines, however, does not 
change the best-fit intensities of the Fe K$\alpha$ lines.
The best-fit intensities of the three iron lines and the 5--10 keV band are shown in table \ref{tab:Tab4ListOfIntensity}.
The profiles of the Fe \emissiontype{XXV} K$\alpha$ line intensity along the Galactic longitude 
($b_*=\timeform{0D}$) and latitude ($l_*=\timeform{-0D.114}$) are shown in figure \ref{fig:FeHeKaProfile} as examples. 

We see global symmetry between the intensities in the positive and negative $l_*$ (figure 3a) and  $b_*$ (figure 3b),
except for the region of $|l_*|=\timeform{0D.1}$--$\timeform{0D.3}$ in figure 3a (see section 4.3).

\begin{table*}
	\caption{List of observed intensities.$^*$}
	\label{tab:Tab4ListOfIntensity}
	\begin{center}
    	\begin{tabular}{rrrrrcrcrccc}
     	\hline
      	\hline
	\multicolumn{4}{c}{Galactic coordinate} & \multicolumn{7}{c}{Observed intensity$^\dagger$}  & Flag$^\ddagger$ \\ 
	$l~(\timeform{D})$ &   $b~(\timeform{D})$ & $l_*~(\timeform{D})$ &   $b_*~(\timeform{D})$ &\multicolumn{2}{c}{Fe \emissiontype{I} K$\alpha$}&\multicolumn{2}{c}{Fe \emissiontype{XXV} K$\alpha$} & \multicolumn{2}{c}{Fe \emissiontype{XXVI} K$\alpha$} & 5--10 keV  & \\
	\hline
	-0.156 & -0.046 & -0.100 & 0.000 & 7.31 &(6.83--7.72)& 18.4 &(17.7--18.9)& 7.97 &(7.44--8.42)& 20.7 & $l$\\ 
	-0.056 & -0.046 & 0.000 & 0.000 & 12.4 &(11.5--13.3)& 45.5 &(44.3--46.7)& 13.4 &(12.9--14.3)& 44.3 & $l$\\ 
	0.044 & -0.046 & 0.100 & 0.000 & 35.2 &(34.8--35.7)& 24.3 &(23.9--24.7)& 9.49 &(9.14--9.85)& 39.0 & $l$\\
	0.500 & -0.146 & 0.556 & -0.100 & 8.35 &(8.05--8.65)& 8.05 &(7.75--8.36)& 3.14 &(2.88--3.41)& 11.3 & $b$\\ 
	0.500 & -0.046 & 0.556 & 0.000 & 9.95 &(9.71--10.3)& 10.2 &(9.95--10.5)& 3.25 &(3.01--3.52)& 13.7 & $b$\\ 
	0.500 & 0.054 & 0.556 & 0.100 & 4.79 &(4.51--5.02)& 9.06 &(8.68--9.32)& 2.28 &(2.02--2.53)& 9.83 & $b$\\  
	0.000 & -2.000 & 0.056 & -1.954 & 0.04 &(0.00--0.12)& 0.64 &(0.55--0.74)& 0.27 &(0.18--0.36)& 1.08 & $o$\\ 
	\hline
     	\end{tabular}   
    	\end{center}
$^*$This table is available in its entirety as a machine-readable form in the online journal (see the PASJ web site).
A portion is shown here for guidance regarding its form and contents. \\
$^\dagger$The interstellar absorption is not corrected. 
The units are $10^{-7}$ photons s$^{-1}$ cm$^{-2}$ arcmin$^{-2}$ and $10^{-14}$ erg s$^{-1}$ cm$^{-2}$ arcmin$^{-2}$, 
for the Fe K$\alpha$ lines and the 5--10 keV band, respectively. Parentheses show the 90\% error ranges.\\ 
$^\ddagger$The flags $l$ , $b$  and $o$ mean that the data are obtained along the Galactic longitude (figure \ref{fig:Fig1}a),
 the latitude (figure \ref{fig:Fig1}b) and from  $(l_*, b_*)=(\timeform{0D.056}$, $\timeform{-1D.954})$, respectively.
\end{table*}

\section {Discussion}\label{ch:discussion}
\subsection{The distributions of the Fe \emissiontype{XXV} K$\alpha$ line and the stellar mass} 
Using the intensity profile, origin of the Fe \emissiontype{XXV} 
K$\alpha$ line in the GCDX is discussed in comparison with that 
in the GRXE.
We fitted the profile of the Fe \emissiontype{XXV} K$\alpha$ line with a stellar mass distribution (SMD) model
which was originally compiled  by \citet{Mu06}, based on the NIR observations \citep{La02, Ke91}. 
The SMD model consists of four components of the nuclear stellar cluster (NSC), 
the nuclear stellar disk (NSD), the Galactic bulge (GB) and the Galactic disk (GD). 
The details of the SMD model are described in Appendix.

We made simultaneous fit for the longitudinal distribution (figure 1a) and for many latitudinal distributions (figure 1b). 
In this fitting, we included the results of  \citet{Ko07b} ($|l_*|<\timeform{0D.4}$), \citet{Ya09} $(|l_*| > \timeform{8D})$ and 
the result of  $(l_*, b_*) =(\timeform{0D.056}, \timeform{-1D.954})$. These data
were added in the longitudinal plot in figures 4a and c, except for those of \citet{Ya09} of $|b| > \timeform{0.15D}$. 
The additional data points in figures 4a and c  are not exactly on $b_*=\timeform{0D}$. 
However, in the simultaneous fitting, the real coordinate values were taken into account.

\subsubsection{The GRXE region}\label{ch:GRXEregion} 
The intensity distribution of the Fe K$\alpha$ line along the Galactic plane has been already reported 
by \citet{Ya93} and \citet{Re06}. Since \citet{Re06} found that the Fe K$\alpha$ line intensity 
profile agrees well with the NIR profile in the GRXE region using the RXTE data, 
we first tried to fit the data of $|l_*|>\timeform{1D.5}$ regions and $|b_*|>\timeform{0D.5}$ 
regions simultaneously with the SMD model.
We assumed that the mass emissivities (see equation A7) 
for each component are the same, i.e. $\varepsilon_{\rm NSC}= \varepsilon_{\rm NSD}
=\varepsilon_{\rm GD}=\varepsilon_{\rm GB}=\varepsilon$.  
Thus the free parameter was only one, $\varepsilon$ or the normalization. 
The best-fit parameter was $\varepsilon= (5.85\pm0.03) \times 10^{33}$ photons s$^{-1}$ $M_{\solar}^{-1}$.
The fitting result for the longitudinal distribution is given in figure 4a, 
while that of the latitude  at $l_*=\timeform{-0D.114}$ is given in figure 4b 
as a typical example.  

As is shown in figure 4a, we obtained a nice fit in the GRXE data of 
$|l_*| > \timeform{1D.5}$, in which the GD and partly GB components 
mainly contribute in the SMD model. 
Figure 4b also shows reasonable fit in the region of $|b_*| > \timeform{0D.6}$, 
where the GD and GB components contribute equally.  

We checked the consistency between the Suzaku results (figure 4) and those of \citet{Re06} (figure 3 in their paper). 
The data in the GRXE region of $|l_*|=\timeform{10D}$--$\timeform{20D}$ were compared
since the limited spatial resolution of RXTE made it impossible to resolve the detailed structure around the 
Galactic center (GCDX).
The result of figure 4a is purely for  Fe \emissiontype{XXV} K$\alpha$ line and 
the Fe \emissiontype{XXV}~K$\alpha$ line intensity at $|l_*|=\timeform{10D}$--$\timeform{20D}$ is 
about $1\times10^{-7}$ photons cm$^{-2}$ s$^{-1}$ arcmin$^{-2}$, while that of \citet{Re06} is the sum of 
Fe \emissiontype{I}, \emissiontype{XXV} and \emissiontype{XXVI} K$\alpha$ 
lines of about $4.5 \times 10^{-4}$ photons cm$^{-2}$ s$^{-1}$ deg
$^{-2}$  (hence $1.3\times10^{-7}$ photons cm$^{-2}$ s$^{-1}$ arcmin$^{-2}$).  
The Fe K$\alpha$ line complex in the RXTE data was not separated into the three lines because of the limited energy  resolution.
From table 1 of \citet{Eb08} and also from figure 3 of \citet{Ya09}, 
we estimated the mean flux ratio of  Fe \emissiontype{I}, \emissiontype{XXV} and 
\emissiontype{XXVI} K$\alpha$ lines as $0.2:1.0:0.2$.  
Thus the Fe \emissiontype{XXV} K$\alpha$ flux is about 70\% of the total flux of Fe K$\alpha$ line complex.  
Taking this flux ratio into account, our results are consistent with that of \citet{Re06}. 
We, therefore, do not discuss origin of Fe \emissiontype{XXV} K$\alpha$ line in the GRXE further than \citet{Re06}.

\subsubsection{The combined regions of the GCDX and GRXE}
In the previous section \ref{ch:GRXEregion}, we found that 
Fe \emissiontype{XXV} K$\alpha$ profile in the GCDX shows clear 
excess above the SMD model which gave a nice fit to that in  the GRXE. 
As is shown in figure 4a, this excess in the GCDX is confined in the region of the NSC and NSD.  
Therefore we fitted the whole data of Fe \emissiontype{XXV} K$\alpha$ line (the GCDX+GRXE region) simultaneously, 
fixing $\varepsilon_{\rm GD}=\varepsilon_{\rm GB}= 5.85 \times 10^{33}$ 
photons s$^{-1}$ $M_{\solar}^{-1}$ and two free parameters of  
$\varepsilon_{\rm NSC}$ and $\varepsilon_{\rm NSD}$.  
The best-fit results are given in figures 4c and d, which 
show globally a good fit with the enhanced  parameter 
values of $\varepsilon_{\rm NSC}$ and $\varepsilon_{\rm NSD}$.  
The best-fit $\varepsilon_{\rm NSC}$ and $\varepsilon_{\rm NSD}$ 
are listed in table 4. 
As is found from table 4, we needed $3.8\pm0.3$ and $19\pm6$ times higher 
emissivities in the NCD and NSC region (GCDX) 
than that in the GD+GB region (GRXE). 

The fitting in which the mass emissivities of the NSD and NSC ($\varepsilon_{\rm NSD}$
and $\varepsilon_{\rm NSC}$) are fixed to the same value of the GD and GB, 
but the stellar masses of the NSC and NSD  ($\rho_{\rm c}$ and $\rho_{\rm d}$) 
are free parameters (see equation \ref{eq:nsc} and \ref{eq:nsd})  gives that
 respective stellar masses of  the NSC and the NSD 
are $3.8\pm0.3$ and $19\pm6$ times larger than those obtained with the NIR 
observations \citep{La02}, normalized to the Galactic ridge. 

In section 4.3, we will discuss the origin of  Fe \emissiontype{XXV} K$\alpha$ line in the 
GCDX based on these results.

\subsection{Systematic error for Fe \emissiontype{XXV} K$\alpha$ line profile and the SMD model}
Since the statistics is limited, present spectral analysis in the small region 
(\timeform{0.1D}$\times$\timeform{0.2D}) 
cannot confine the interstellar absorption ($N_{\rm H}$) within reasonable errors.
We therefore did not correct the absorption effect for the flux of Fe \emissiontype{XXV}~K$\alpha$.
The largest interstellar absorption is $N_{\rm H} \sim 6 \times10^{22}$ cm$^{-2}$ near the 
Galactic center \citep{Pr95}. Using the cross section of the 
photoelectric absorption of \citet{Mo83}, we estimated that the flux of Fe \emissiontype{XXV} K$\alpha$ 
is reduced only by 9\% in the largest case of $N_{\rm H}=6 \times10^{22}$ cm$^{-2}$. 
Since the largest reduction is expected in the GCDX region, 
the excess discussed in section 4.1.2 becomes even larger by about 10\%,  
if we correct the interstellar absorption.
A larger error is in the making of the SMD model,  which is estimated to be about 50\% \citep{Mu06} (see Appendix). 
Taking these errors into account, our conclusion of the large excess of Fe \emissiontype{XXV}~K$\alpha$ line at the GCDX to the GRXE in 
the best-fit SMD model is still valid, and hence no essential change of the discussion is required.

\subsection{Origin of Fe \emissiontype{XXV} K$\alpha$ line in the GCDX}
We found that  the mass emissivity of Fe \emissiontype{XXV} K$\alpha$ line
in the GCDX region (NSC and NSD) 
shows excess of $3.8\pm0.3$ (NSC) and $19\pm6$ (NSD) times larger than 
that in the GRXE region (GD and GB), 
or the stellar masses in the NSC and NSD regions are $3.8\pm0.3$ and 
$19\pm6$ times larger than those estimated  with the NIR observations.
Integrating over the whole region of the NSC and NSD, we obtain that
the total stellar mass in the NSC and NSD are about 
$1.4 \times 10^9$~$M_{\solar}$ for the former case, which is
fully consistent with the kinematic mass distribution (see figure 14 of \cite{La02}).
However the latter case  requires total stellar mass within the NSC and NSD to be
$5.3 \times 10^9$~$M_{\solar}$, which is unacceptably large. We hence ignore the latter
possibility, and discuss based on the former possibility.

Origin of Fe \emissiontype{XXV} K$\alpha$ line in the GCDX therefore cannot be the same as that in the GRXE.
\citet{Ya09} found that the line intensity ratio of Fe \emissiontype{XXVI} K$\alpha$ to Fe \emissiontype{XXV} K$\alpha$ 
in $|l_*| < \timeform{0D.3}$ is higher than that in $|l_*| > \timeform{8D}$, which also supports  that  
origin of Fe \emissiontype{XXV} K$\alpha$ line in the GCDX may be different from that in the GRXE.
\citet{Re09} resolved more than 80\% of the Fe K$\alpha$ flux at $(l,\ b)=(\timeform{0D.1}, \timeform{-1D.4})$
into point sources. 
Although this fact supports that origin of Fe \emissiontype{XXV} K$\alpha$ line in the GRXE is due to point sources, that in the
GCDX is not yet answered. As is shown by the arrows in figures 4b and 4d,  Fe \emissiontype{XXV}~K$\alpha$ from
this region is due to the sum of nearly equal contribution of the GD and GB (the GRXE
components) and not the GCDX components (the NSC and NSD).
 
If we assume that origin of Fe \emissiontype{XXV} K$\alpha$ line in the GCDX is an ensemble of point sources 
with the same populations in the GRXE region, the
iron abundances in the GCDX should be 
$3.8(\pm0.3$)--$19(\pm6$) 
times larger than that in the GRXE.
This is unlikely because the abundance of iron has been reported at 
most two times of solar in the GCDX region (e.g. \cite{Ko07b,No10}). 

Another possibility is that the point source population in the GCDX
is different from that in the GRXE. 
We see the Fe \emissiontype{XXV} K$\alpha$ intensity
excess in the positive $l_*$ side (east) compared to 
the negative side (west) at $|l_*|=\timeform{0D.1}
$--$\timeform{0D.3}$ (see figure 3a).
This excess is globally associated to the clusters 
of high mass stars such as the Arches and Quintuplet clusters.
\citet{Mu09} also reported that the number density of X-ray stars in these regions 
exceeds that in the surroundings.
They proposed that the excess is due to massive stars associated to the 
high activity of star formation in this region.
The most probable massive stars are WR or OB stars.
These stars often show moderately large
equivalent width of Fe \emissiontype{XXV} K$\alpha$ ($EW_{6.7}$) similar to the GCDX (e.g. \cite{Sk07, Hy08}). 
These massive stars might contribute to not only the Arches and Quintuplet
clusters but also the entire GCDX.
These stars are relatively bright with the typical X-ray luminosity of  $\sim 10^{34}$ erg s$^{-1}$ \citep{Po02, Wa06}. 
The deep Chandra observation revealed that point sources with the luminosity larger than 
$10^{32}$ erg s$^{-1}$  contribute less than 10\% of the GCDX (see figure 6 of \cite{Re07GC}).  
We therefore concluded that massive stars do not mainly contribute to Fe \emissiontype{XXV} K$\alpha$  line in the GCDX.

On the other hand, it may be conceivable that faint X-ray point sources with the flux
below the current detection limits largely contribute to the GCDX.
The most probable candidate sources for the Fe \emissiontype{XXV} K$\alpha$ line are
cataclysmic variables (CVs), because their spectra are represented by high temperature plasma
with large $EW_{6.7}$ and are associated with the 6.4 and 7.0 keV lines, very similar to those 
of the GCDX in the Fe K$\alpha$ line band.
The mean $EW_{6.7}$ for X-ray bright CVs in the luminosity range of
$10^{30-33}$ erg s$^{-1}$ is about 200 eV \citep{Ez99, Ra06},
while those  of the GCDX and GRXE are 450--550 eV \citep{Ya09}. 
Thus even for the GRXE, the faint (unresolved X-ray) CVs should have, 
at least, two times larger $EW_{6.7}$ than the bright CVs. 
In addition, the number density of such (X-ray) faint CVs relative to
near infrared stars should be $3.8(\pm0.3$)--$19(\pm6$) times larger than that in the GRXE 
to explain the observed Fe \emissiontype{XXV} line excess in the GCDX.
This possibility does not violate
the total mass constraint of the NSC and NSD ($\sim1.4 \times 10^9$ $M_{\solar}$), because the mass contribution
of CVs to that of the SMD (mass of infrared stars) is only a small fraction
(see e.g., \cite{Pr07} and references there in).
For relatively bright CVs, however, such number density excess in the GCDX
region has not been reported \citep{Re08}. 
For faint CVs,  studies for the spatial profile have not been done well.
In order to settle these potential problems in the point source origin, deep observations,  
with lower detection limit than the current value by orders of more than 1--2 are required.

An alternative possibility is that the majority of the GCDX (NSC and NSD) comes from diffuse optically thin thermal 
plasma as already proposed  by \citet{Ko07b}. 
Although the total flux of the Fe \emissiontype{XXV} K$\alpha$ line in the NSC and NSD regions is hardly explained by point sources, 
the spatial distributions are similar to the shapes of the NSC and NSD. 
The diffuse plasma, therefore, should be closely related to the ensemble of point sources.  
One possibility is that the plasma is made by multiple supernova explosions. 
This may be conceivable because the NSC and NSD regions are distinguished from the other regions (GB and GD) in respect to supernova activities; i.e., very high density of  molecular gas and on-going star formation.

\begin{figure*}[htbp]
	\begin{center}
		\begin{center}
		(a)\FigureFile(72mm,54mm){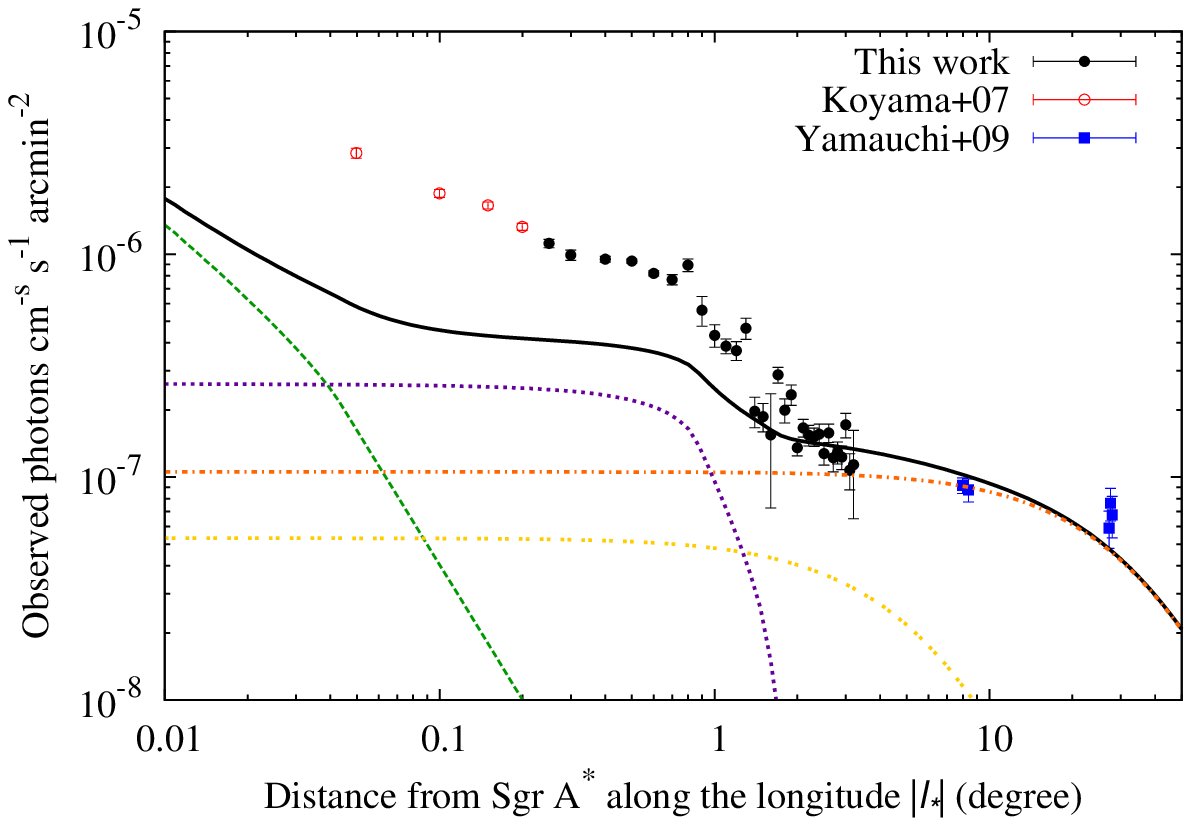}
		(b)\FigureFile(72mm,54mm){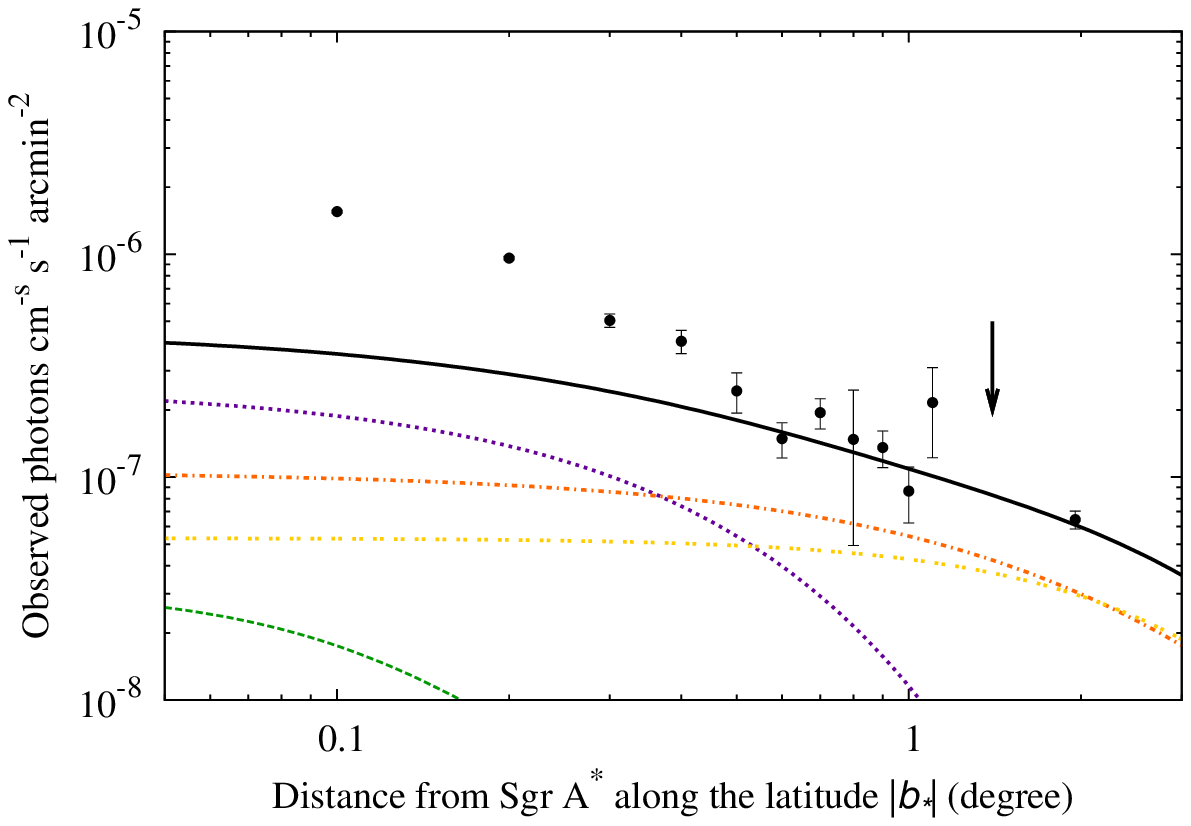}
		(c)\FigureFile(72mm,54mm){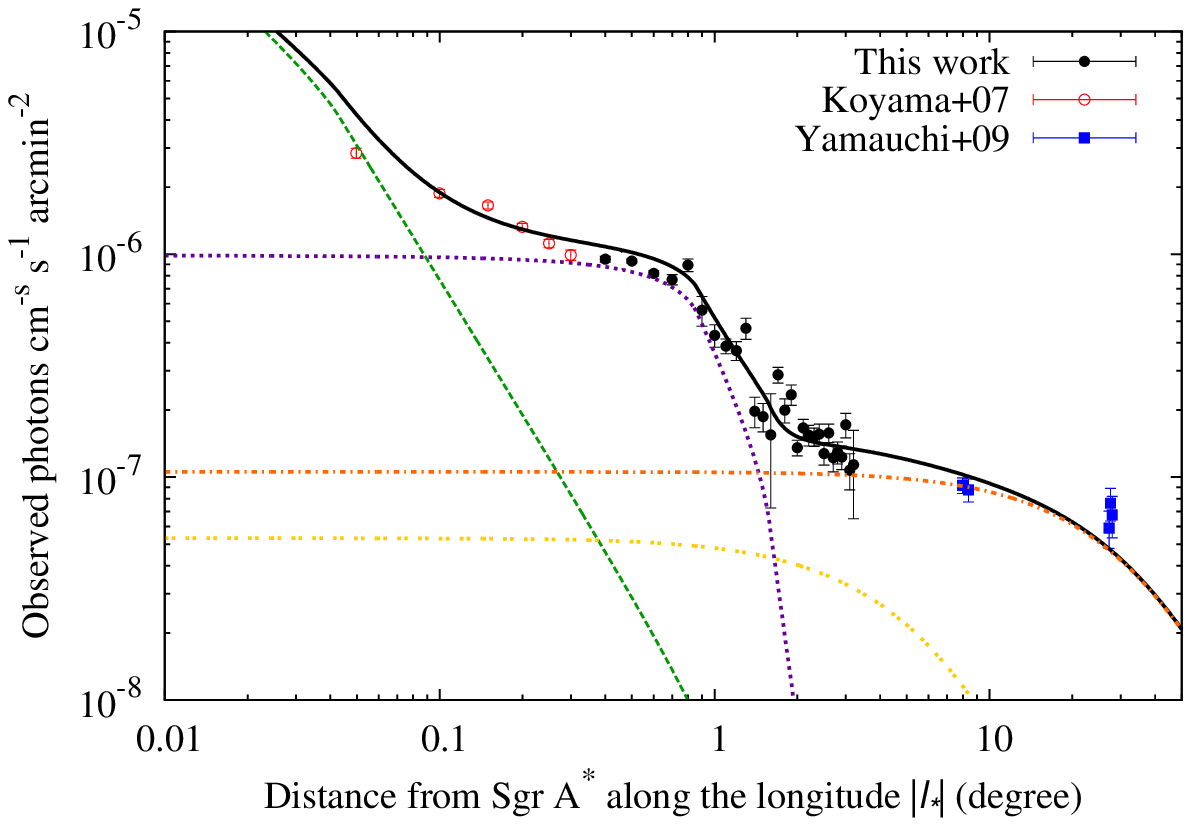}
		(d)\FigureFile(72mm,54mm){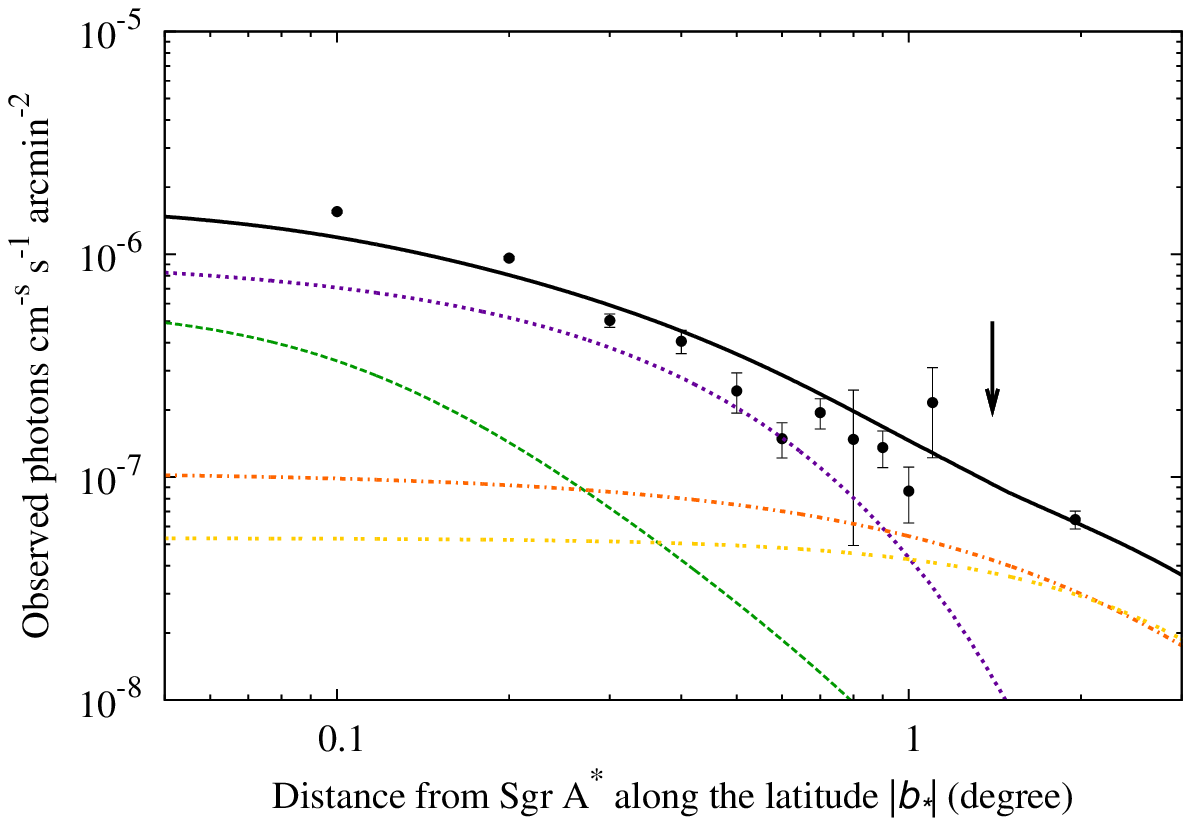}
	\end{center}
	\caption{(a) The Fe \emissiontype{XXV} K$\alpha$ line intensity distribution along the Galactic plane as a function of $l_*$ (distance from Sgr A$^*$). For simplicity, only negative $l_*$ region is shown, except for those at  $|l_*| > \timeform{8D.0}$ because the model in these region (Galactic disk) is symmetric with respect to the positive and negative $l_*$.  The green, purple, orange and yellow 
		lines are the best-fit SMD model of the nuclear stellar cluster (NSC), nuclear stellar disk (NSD), Galactic disk (GD) 
		and Galactic bulge (GB), respectively.  The black solid line is the sum of the four components of NSC, NSD, GD and GB.  
		The emissivities of these components are fixed to be the same. The data of $|l_*|>\timeform{1D.5}$ and $|b_*|>\timeform{0D.5}$ 
regions are used simultaneously in the fitting.\newline 
                 (b) The same as the left panel (a), but those along the Galactic longitude at $l_*=\timeform{-0D.114}$ 
                as a function of $b_*$ (distance from the Galactic plane). 
		The arrow shows the position of $b=\timeform{-1D.4} $ which is almost the same region as \citet{Re09}. 
		For simplicity, only negative $b_*$ region is  shown.\newline
 		(c) The same as the upper panel (a) but with the best-fit model in the whole region (GRXE and GCDX) with free parameters of 
		the emissivities (or stellar mass densities) of NSC and NSD.\newline
		(d) The same as the upper panel (b) but with the best-fit model same as the panel (c).\newline
		}
		\label{fig:SMM_fit}
		\end{center}
\end{figure*}	

\begin{table}
 \caption{Best-fit result with the stellar mass distribution model.}
  \label{tab:SMM_fit}
  \begin{center}
    \begin{tabular}{cc}
     \hline
      \hline
	Component& Mass emissivity of \\
	&the Fe \emissiontype{XXV} K$\alpha$ line $\varepsilon$  ${}^*$\\     
      \hline 
      Nuclear stellar cluster & 111 $\pm$ 37\\
      Nuclear stellar disk  & 22 $\pm$ 2 \\
      Galactic disk$^\dagger$ & 5.85 (fixed) \\
        \hline
      \end{tabular}
    \end{center}
   ${}^*$ See equation \ref{eq:integral2}. The units are $10^{33}$ photons s$^{-1}$ $M_{\solar}^{-1}$. 
	${}^\dagger$ Mass emissivities of the Galactic bulge and the Galactic disk are fixed to the best-fit 
        of $|l_*|>\timeform{1D.5}$ and $|b_*|>\timeform{0D.5}$ regions. 
\end{table}

\section{Summary}
\begin{enumerate}

\item We have obtained  intensity profiles of the Fe~\emissiontype{I}, 
Fe~\emissiontype{XVV}, and Fe~\emissiontype{XXVI} K$\alpha$ lines in 
the Galactic center region of $\timeform{-3D}< l < \timeform{2D}$ and $\timeform{-2D} < b <\timeform{1D}$.

\item The intensity profile of Fe \emissiontype{XXV} K$\alpha$ is nicely fitted with the SMD model in the GRXE region 
($|l_*|>\timeform{1D.5}$ or $|b_*|>\timeform{0D.6}$), while that in the GCDX ($|l_*|<\timeform{1D.5}$ and $|b_*|<\timeform{0D.6}$) 
shows $3.8(\pm0.3)$--$19(\pm6)$ times excess over the best-fit SMD model to the profile of the GRXE region.

\end{enumerate}

\appendix
\section*{Stellar Mass Distribution Model of the Milky Way Galaxy}\label{ch:SMM_eq}
This appendix describes the details of the three-dimensional stellar mass distribution (SMD) 
and actual two-dimensional fitting model (SMD model) in section 4. The SMD model was originally compiled by \citet{Mu06} 
using the results of  \citet{La02} and \citet{Ke91} based on near infrared (NIR) observations with COBE, IRAS and IRT \citep{Bo92, Cl80, Ko82}. 
The major emission sources at the NIR wavelength (1--7 $\mu$m) are main-sequence stars and thus the NIR emission is a good tracer of the stellar mass. 
The stellar mass model consists of the components of the nuclear stellar cluster (NSC), the nuclear stellar disk (NSD), Galactic bulge (GB) 
and Galactic disk (GD).  In order to describe the three-dimensional stellar mass distribution of these components, we defined the X-Y-Z coordinate, 
where Sgr A$^*$ is located at the origin $(x, y, z)=(0, 0, 0)$, and the Galactic plane is the X-Y plane. The sun is assumed 
to be at (8500, 0, 16) in units of pc.

The stellar mass density model $\rho_*$ (${M_{\solar}}~{\rm pc}^{-3}$) is 
$
\rho_*=\rho_{\rm NSC} + \rho_{\rm NSD}+ \rho_{\rm GD} + \rho_{\rm GB}. 
$
The details of each component are described below. 

The NSC component  is 
\begin{equation}
\rho_{\rm NSC} = \frac{\rho_{\rm c}}{1+(r/r_{\rm c})^{n_{\rm c}}}, \label{eq:nsc}
\end{equation}
where $r=(x^2+y^2+z^2)^{1/2}$ is the distance from the coordinate origin (0, 0, 0), Sgr A$^*$.
The parameters are $r_{\rm c}=0.22$~pc, $\rho_{\rm c}=3.3\times10^6~M_{\solar}~{\rm pc^{-3}}$ ($r < 6$~pc), 
$9.0\times10^7~M_{\solar}~{\rm pc^{-3}}$ ($ 6 \leqq r < $ 200 pc) and 
$n_{\rm c}=2$ $(r < 6$~pc), 3 ($ 6 \leqq r < $ 200 pc). 

The NSD and GD components are respectively given as, 
\begin{equation}
\rho_{\rm NSD} =\rho_{\rm d} \left(\frac{r}{r_{\rm d}}\right)^{-{n_d}}\exp\left(-\frac{|z|}{z_{\rm d}}\right), \label{eq:nsd}
\end{equation}
and
\begin{eqnarray}
\rho_{\rm GD} = \rho_{\rm gd} \exp\left(-\frac{r}{r_{\rm gd}}\right)\exp\left(-\frac{|z|}{z_{\rm gd}}\right),
\end{eqnarray}
where $r=(x^2+y^2)^{1/2}$ and $z$ are the distance from Sgr A* in the Galactic plane and the height from the Galactic plane, respectively.
The parameters for the NSD are $r_{\rm d}=1$~pc, $z_{\rm d}=45$~pc, 
and $n_{\rm d}=0.1$ ($r < 120$~pc), 
3.5 ($120 \leqq r < $ 220 pc), 
10 ($220 \leqq r < $ 2000 pc). 
We took $\rho_{\rm d}=300~M_{\solar}~{\rm pc^{-3}}$ for $r < 120$~pc, 
and adjusted $\rho_{\rm d}$ to make the function continuous at $r$ =120 and 220 pc.
The parameters for the GD are $\rho_{\rm gd}=5~M_{\solar}~{\rm pc^{-3}}$, $r_{\rm gd}=2700$ pc and $z_{\rm gd}$=200 pc.

The GB component is 
modeled as a triaxial ``generalized ellipsoidal'' bar \citep{At90, Fr98}.  Since the major axis of this ``ellipsoid'' is tilted from any 
of the X-, Y- and Z-axis, we defined a new coordinate ${\rm X^\prime}$-${\rm Y^\prime}$-${\rm Z^\prime}$ by rotating the  X-Y-Z coordinate along 
the Z-axis by $\timeform{15D}$ to the east and along the Y-axis by $\timeform{1D}$ to the north. 
In the new coordinate ${\rm X^\prime}$-${\rm Y^\prime}$-${\rm Z^\prime}$, the GB is given by;
\begin{eqnarray}
\rho_{\rm GB} = \rho_{\rm gb}~\exp(-r_{\rm s}), \\
r_{\rm s} = \left[r^{c_{\parallel}}_{\perp} + \left(\frac{|z^\prime|}{a_z}\right)^{c_{\parallel}} \right]^{1/c_{\parallel}},\\
r_{\perp} = \left[\left(\frac{|x^\prime|}{a_x}\right)^{c_{\perp}} + \left(\frac{|y^\prime|}{a_y}\right)^{c_{\perp}} \right]^{1/c_{\perp}}.
\end{eqnarray}
The parameters are  $\rho_{\rm gb}=8~M_{\solar}~{\rm pc^{-3}}$, $a_x$=1100 pc, $a_y$=360 pc, $a_z$=220 pc, $c_{\parallel}=3.2$ and $c_{\perp}=1.6$.

For the three-dimensional mass distribution function $\rho(x, y, z)$ 
of the components NSC (A1) , NSD (A2),  GD (A3) and GB (A4--6), we 
defined the mass emissivity of Fe\emissiontype{XXV} K$\alpha$ 
line $\varepsilon$ (in units of photons s$^{-1}$ $M_{\solar}^{-1}$) 
to be the ratio of Fe \emissiontype{XXV} K$\alpha$ photons emitted 
from a unit stellar (infrared stars) mass per second.
The number of Fe \emissiontype{XXV} K$\alpha$ photons emitted per second from a small volume in the solid angle 
$\Omega$ (a rectangle of $\timeform{0D.1} \times \timeform{0D.2}$)
and depth $\Delta s$ at the distance $s$ from the Sun toward ($l_*$ ,$b_*$) direction is given as 
$\Omega s^2 \Delta s \varepsilon \rho$.
The observed flux $\Delta F(l_* ,b_*)$  from the small volume should be  
$\Omega s^2\Delta s  \varepsilon \rho /4\pi s^2= \Omega \varepsilon \rho \Delta s / 4\pi$.
Then the observed flux of Fe \emissiontype{XXV} K$\alpha$ is given by
the line-of-sight integration along $s$,
\begin{eqnarray}\label{eq:integral2}
F(l_* ,b_*)= \frac{\Omega}{4\pi}\int_{0}^{\infty} (\varepsilon_{\rm NSC}~\rho_{\rm NSC} + \varepsilon_{\rm NSD}~\rho_{\rm NSD} \\ \nonumber
+\varepsilon_{\rm GD}~\rho_{\rm GD} +  \varepsilon_{\rm GB}~\rho_{\rm GB}) ds. 
\end{eqnarray}
This equation is used for the fitting of observed intensity profile of  Fe \emissiontype{XXV} K$\alpha$ line (figure 4 in the text). The overall uncertainty in building of this model is estimated to be about 50\% (\cite{Mu06}).

\bigskip

The authors thank all of the Suzaku team members, especially 
M. Sawada, T. Yuasa, and  K. Makishima for their comments and useful information.
This work is supported by the Grant-in-Aid for the Global COE Program ``The Next Generation 
of Physics, Spun from Universality and Emergence'' and Challenging Exploratory Research (KK) 
from the Ministry of Education, Culture, Sports, Science and Technology (MEXT) of Japan.
HU is supported by Japan Society for the Promotion of Science (JSPS) Research Fellowship for Young Scientists.
HM is supported by Grant-in-Aid for Scientific Research (B)
(22340046) from JSPS.

\end{document}